\documentclass{article}
\usepackage{spconf,amsmath,epsfig}
\usepackage{color}
\usepackage{subfigure}

\begin{document}\sloppy

% Example definitions.
% --------------------
\def\x{{\mathbf x}}
\def\L{{\cal L}}

% Title.
% ------
\title{HEVC Inter Coding using Deep Recurrent Neural Networks and Artificial Reference Pictures}
%
% Single address.
% ---------------
\name{Felix Haub, Thorsten Laude\thanks{Corresponding author: Thorsten Laude (laude@tnt.uni-hannover.de)} and J\"orn Ostermann}
\address{Leibniz University Hannover, Institut f\"ur Informationsverarbeitung, \\Appelstr. 9a, 30167 Hannover, Germany}

\maketitle

\begin{abstract}

The efficiency of motion compensated prediction in modern video codecs highly depends on the available reference pictures.
Occlusions and non-linear motion pose challenges for the motion compensation and often result in high bit rates for the prediction error.
We propose the generation of artificial reference pictures using deep recurrent neural networks.
Conceptually, a reference picture at the time instance of the currently coded picture is generated from previously reconstructed conventional reference pictures.
Based on these artificial reference pictures, we propose a complete coding pipeline based on HEVC.
By using the artificial reference pictures for motion compensated prediction, average BD-rate gains of 1.5\% over HEVC are achieved.

\end{abstract}
\begin{keywords}
Video Coding, HEVC, Deep Learning, RNN, Motion Compensation
\end{keywords}
\section{Introduction}
\label{sec:intro}
High Efficiency Video Coding (HEVC) was technically finalized in January 2013 and constitutes the standardized state-of-the-art for video coding since then \cite{Sullivan2012}. 
As a joint effort of the Joint Collaborative Team on Video Coding of ISO/IEC and ITU-T, it was published as MPEG-H Part 2 and \mbox{H.265}, respectively.
Compared to its predecessor standard Advanced Video Coding (AVC), HEVC enables a 40--60\% bit-rate reduction while maintaining a comparable visual quality \cite{Hanhart2012,DeCock2016}.
The consistently high desire for improved coding efficiency motivated the continued research for compression algorithms beyond HEVC, for example JEM or AV1 \cite{Laude2018}.

All of the named video codecs share the same fundamental working principle: block-based hybrid video coding.
It consists in the combination of a prediction with transform coding for the prediction error.
The prediction methods can be distinguished into intra and inter coding.
Intra coding relies on previously coded parts of the current picture to predict a new block within this picture.
Inter coding additionally utilizes temporal redundancy between consecutive pictures to improve the prediction.
Conceptually, previously reconstructed pictures are stored in a reference picture buffer and used to make a prediction for the currently coded block via motion compensated prediction.
The quality of motion compensated prediction highly depends on the available reference pictures.
Furthermore, the better motion compensation performs, the lower the bit rate for the prediction error gets.

It is worth noting that due to the motion compensation, the quality of the reference pictures does not necessarily correlate with the pixel-wise fidelity between the current picture and the reference pictures.
For example, a reference picture which is a translationally shifted version of the current picture would be a good prediction reference despite the low pixel-wise fidelity between these pictures.
More problematic are complex motions or occlusions which cannot be handled by the motion model of the video codec.

In this paper, we use a deep learning-based \cite{Sze2017} approach to overcome this limitation.
Conceptually, we process existing reference pictures from the buffer (which are referred to as \textit{conventional reference pictures} in the following) with a recurrent neural network to generate a new \textit{artificial reference picture}. 
This artificial reference picture is then additionally used for motion compensated prediction.
The underlying hypothesis of our work is that the artificial reference picture enables a better prediction which in turn results in a smaller prediction error and a lower bit rate.
Our main contributions in this paper are:
\begin{itemize}
    \item Generation of artificial reference pictures using a recurrent neural network
    \item Complete coding pipeline with the neural network integrated in the video codec HEVC
\end{itemize}
    
The remainder of this paper is organized as follows:
In Section~\ref{sec:related_works}, we discuss the closest related works and highlight the distinguishing features of our proposed method.
Our method for deep learning-based reference picture generation is presented in Section~\ref{sec:method}.
The experimental results are presented and discussed in Section~\ref{sec:evaluation}.
In Section~\ref{sec:conclusions} we draw our conclusions for this work.

\section{Related Works}
\label{sec:related_works}
In this section, we briefly review the closest related works in the following three categories:
generation of additional reference pictures for improved prediction, usage of deep learning for improving video coding, prediction of future pictures from a sequence of pictures using neural networks.

In the first category, Laude \textit{et al.} generate a new reference picture in the context of scalable video coding where multiple representations of a video (e.g. different qualities or resolutions) are coded jointly \cite{Laude2014a,Laude2014b}. 
Basically, they combine low-frequency information from base layers with high frequency information from enhancement layers.
For this purpose, they apply adaptive Wiener filters to reference pictures of both layers and inter-layer motion compensation to the enhancement layer.
In their approach, the existence of multiple representations of the same video is indispensable.
In contrast to that, our method is applicable to the general case of video coding in which only one representation of a video is coded. 

Works in the second category gained popularity during the last few years as deep learning spread to many new applications.
Video encoding is a very complex task because a comprehensive rate-distortion optimization is required to fully exploit the capability of modern video codecs.
Therefore, neural networks were adopted to approximate the optimal rate-distortion decision (e.g. \cite{Laude2016b}).
In contrast to that we use neural networks for a novel coding algorithm instead of for the control of existing coding modes.
Li \textit{et al.} in \cite{Li2017} and Li \textit{et al.} in \cite{Li2018} propose deep learning-based algorithms for intra coding.
In contrast to that, we use neural networks in the context of inter prediction.

For our method, we use a deep neural network to predict a picture from a sequence of preceding pictures. 
There are a number of related works covering this problem which fall into the third category of related works, amongst them \cite{Liu2017, Mathieu2016, Lotter2017}. 

In \cite{Liu2017}, Liu \textit{et al.} train a convolutional encoder-decoder neural network to calculate the optical flow between two or more pictures. 
Using the optical flow data, the authors synthesize predictions of either in-between pictures (interpolation) or subsequent pictures (extrapolation). 
In \cite{Mathieu2016} and \cite{Lotter2017}, optical flow is not used for picture prediction.

To improve the sharpness of future pictures predicted by a convolutional neural network, Mathieu \textit{et al.} propose a multi-scale neural network, an adversarial training method, and a special loss function in \cite{Mathieu2016}.

In \cite{Lotter2017}, Lotter \textit{et al.} predict future pictures for a sequence of pictures with a recurrent neural network architecture they call \textit{PredNet}. 
The architecture is inspired by the concept of \textit{predictive coding} from the neuroscience literature. 
Predictive coding in this case describes the process of the brain continually making predictions of incoming sensory stimuli which are then compared to the actual incoming sensory stimuli to improve future predictions. 
The authors adopt this principle in that their network performs a prediction for every single picture in the sequence of pictures which is compared to the actual picture at that time instance for improved prediction of the next picture. 
Typically, nine previous pictures were utilized for the prediction of the next picture.
This results in a high-quality prediction when finally a future picture is predicted. 
The PredNet consists of multiple similar modules which make local predictions and only forward the error obtained from this prediction to subsequent modules.

In contrast to the works in the third category we not only use a neural network for future picture prediction, but we also use the prediction to improve motion-compensated prediction in video coding.

\section{Deep Learning-based Reference Picture Generation}
\label{sec:method}
\begin{figure}
    \centering
    \includegraphics[width=\columnwidth]{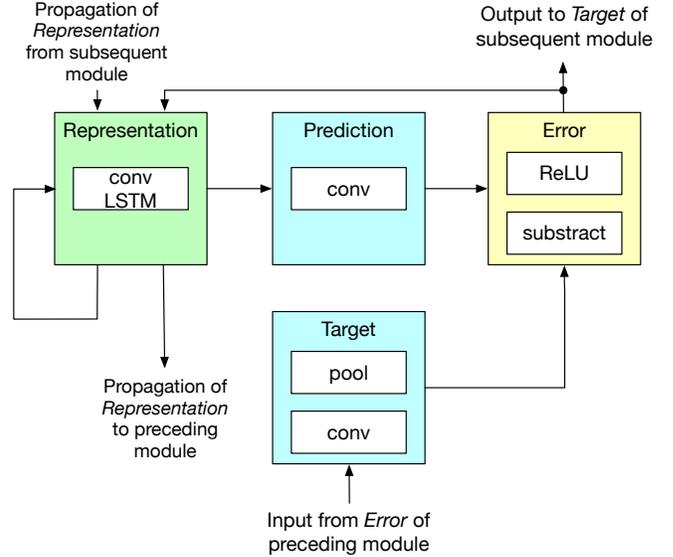}
    \caption{PredNet module structure. The network is formed by four stacked modules. Adopted from \cite{Lotter2017}.}
    \label{fig:prednet}
\end{figure}
\begin{figure}
    \centering
    \includegraphics[width=0.95\columnwidth]{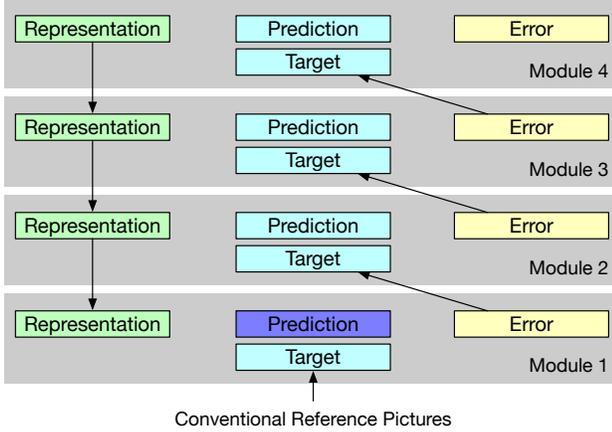}
    \caption{Architecture of the stacked modules (from Fig.~\ref{fig:prednet}). Connections inside modules are not shown for easier readability. The final prediction is generated by the module with slightly darker blue. Based on \cite{Lotter2017}.}
    \label{fig:stacked_modules}
\end{figure}
In this work, we adopt the recurrent neural network architecture from the PredNet model proposed by Lotter \textit{et al.} \cite{Lotter2017} and use it to predict the picture to be coded from its reference pictures. 
For conciseness, the architecture is briefly reviewed in the following.

The network consists of four stacked modules with the same architecture (with differences for the first and last module). 
Every module contains several submodules which are explained in the following as shown in \mbox{Fig. \ref{fig:prednet}}:
A recurrent \textit{Representation} submodule which is a convolutional Long Short-term Memory (LSTM) layer;
a \textit{Prediction} submodule which is a convolutional layer;
a \textit{Target} submodule composed of a convolutional and a pooling layer;
an \textit{Error} submodule which consists of a subtraction between the input and prediction submodules combined with a ReLU activation function.

Four of those modules are used. 
The modules are positioned in a sequence and every module is connected to its preceding and its subsequent module with four connections as shown in Fig. \ref{fig:stacked_modules}. 
We will refer to the modules as Module 1, 2, 3 and 4.

The input to Module 1 is not propagated from a preceding module as there is no preceding module.
The reference pictures are supplied to the Target of Module 1. 
The pictures are supplied, a single one per cycle, one after another in a recurrent process.
Because Module 4 has no subsequent module, there are no connections to a subsequent module.

The first step in the first cycle is the update of the Representation of each module beginning from the back with Module 4.
The Representation of Module 4 is updated with the data of its Error and the Representation is then propagated to Module 3. 
The Representation of Module 3 is in turn updated with the data of its Error and the propagated Representation. 
This process is repeated for every module, finally updating the Representation in Module 1.

The next step in the first cycle is the prediction in each module, this time beginning with Module 1. 
Because no previous pictures are provided, the prediction is empty. 
Meanwhile, the actual reference picture is supplied to the Target of Module 1.
In the Error submodule, the difference (error) between the Prediction and the Target is calculated. 
This error is propagated to the Target of Module 2.
Next, a prediction for the error of Module 1 is generated in Module 2. 
This prediction is then compared to the actual error of Module 1. 
The difference is calculated in the Error submodule and again propagated to Module 3. 
The process is the same for Module 4. After the Error submodule of Module 4 has been updated, the next cycle begins.

The number of cycles is equal to the number of reference pictures plus one.
In the last cycle, the Representations of all four modules are again updated. 
A prediction for the picture to be coded is performed in Module 1. 
This time, no reference picture is supplied to the Target because it is not available for this time instance. 
This prediction is used as an artificial reference picture. 

For further details concerning the PredNet architecture the reader is referred to \cite{Lotter2017}.

In this paper, we differentiate between the terms sequence (whole video) and snippet (five consecutive pictures long portion of a video). 
To train our neural network we used snippets from the \textit{KITTI raw dataset} \cite{Geiger13}. 
The KITTI raw dataset consists of uncompressed traffic recordings which contain a considerable amount of motion.
Using adequate training data is imperative for the performance of neural networks.
Images from many databases like ImageNet suffer from partly severe compression artifacts.
While this is not a major problem for computer vision tasks like image recognition, a problem arises for the regression task of image prediction.
With a compromised database, the network would learn to create compression artifacts.
Therefore, we ensured to use a database with uncompressed data.
Hence, the neural network will not learn any artifacts caused by compression.

From the KITTI raw data set we generated 50000 snippets in the resolution $176 \times 144$ (QCIF). 
During the training process we randomly chose 1000 snippets out of those for every epoch. 
The network was trained for 150 epochs in total.

As model parameters we used $3 \times 3$ convolutions and layer channel sizes of (3, 48, 96, 192) following \cite{Lotter2017}. 
Models were trained with the Adam solver \cite{Kingma2014} using a loss solely computed based on the Error submodule of module one. 
We initially used the default parameter values for Adam, learning rate $\alpha = 0.001$, $\beta_1 = 0.9$, $\beta_2 = 0.999$. 
Additionally, we decreased the learning rate by a factor of 10 halfway through training. 

\begin{figure}
    \centering
    \includegraphics[width=\columnwidth]{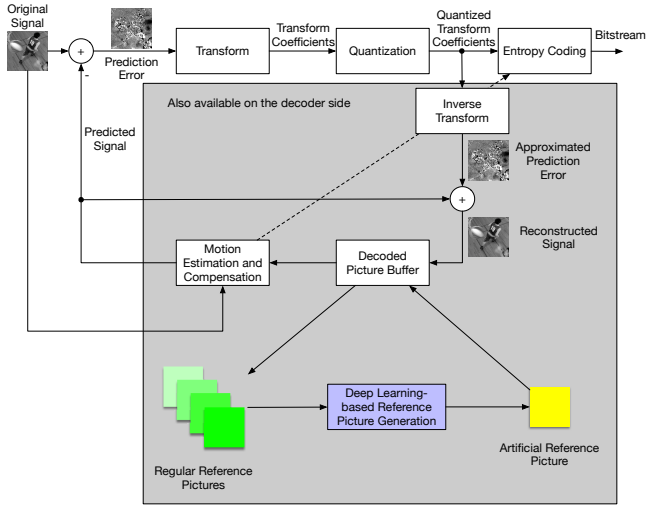}
    \caption{Block diagram of the proposed pipeline. The existing reference pictures are used to generate an artificial reference picture using our deep learning-based reference picture generation.}
    \label{fig:block_diagram}
\end{figure}

We use our trained neural network to generate artificial reference pictures which are used for  the motion-compensated prediction of the HEVC encoding and decoding processes, respectively.
Our modified implementation of HEVC is explained in the following and illustrated in Fig.~\ref{fig:block_diagram}.
Before the encoding or decoding process of every single picture is started all reference pictures from the reference picture list of the picture to be coded are supplied to the neural network.
The neural network generates a prediction of the picture to be coded which can be used as an artificial reference picture.

There are two possible ways to use this artificial reference picture. 
Either the picture can be added to the reference picture list or the picture can replace one of the pictures in the reference picture list.
Since the selection of reference pictures is an encoder choice which is obligatorily signaled as part of the high-level syntax for each slice \cite{Sjoberg2012} -- only exception: Instantaneous Decoder Refresh (IDR) -- this choice does not impose any restriction of the method.

When adding the picture to the reference picture list, it is difficult to measure if the artificial reference picture is a better reference for motion-compensated prediction than the existing reference pictures. 
This is because the additional reference picture might improve motion-compensated prediction only because it is different from the other reference pictures so that it sometimes can give an improved MSE after motion-compensated prediction and not because it is superior to the other reference pictures.
Additionally, the motion-compensated prediction could also be improved by simply adding a conventional reference picture from a not yet considered time instance.

However, if replacing one of the reference pictures with the artificial reference picture leads to an improved coding efficiency during the encoding process then the artificial reference picture is superior to the replaced reference picture. 
For this reason, we chose to replace a reference picture.
Still, our method is not limited to this approach.

We will demonstrate in our evaluation that the reference picture $t_{-4}$ with the highest temporal distance to the currently coded picture $t_{0}$ has the highest MSE and the lowest SSIM compared to $t_{0}$. 
This motivated us to replace reference picture $t_{-4}$ with our artificial reference picture and not any of the other reference pictures.

Other changes to the encoding and decoding process are not necessary for our method because the motion-compensated prediction can utilize the artificial reference picture in the same way as it utilizes the conventional reference pictures.

\section{Evaluation}
\label{sec:evaluation}
\begin{figure*}[ht]

	\centering
	\subfigure{
		\begin{minipage}{0.175\textwidth}
			\centering
			\includegraphics[width=0.95\columnwidth]{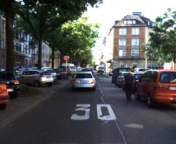}
		\end{minipage}
	}
	\subfigure {
		\begin{minipage}{0.175\textwidth}
			\centering
			\includegraphics[width=0.95\columnwidth]{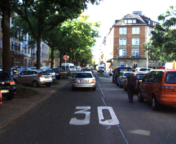}
		\end{minipage}
	}
	\subfigure {
		\begin{minipage}{0.175\textwidth}
			\centering
			\includegraphics[width=0.95\columnwidth]{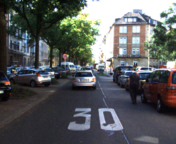}
		\end{minipage}
	}	
	\subfigure {
		\begin{minipage}{0.175\textwidth}
			\centering
			\includegraphics[width=0.95\columnwidth]{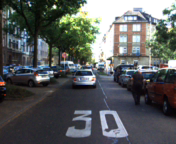}
		\end{minipage}
	}	
	\subfigure {
		\begin{minipage}{0.175\textwidth}
			\centering
			\includegraphics[width=0.95\columnwidth]{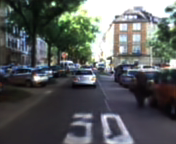}
		\end{minipage}
	}
	\\
	\subfigure{
		\begin{minipage}{0.175\textwidth}
			\centering
			\includegraphics[width=0.95\columnwidth]{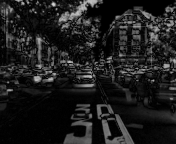}
		\end{minipage}
	}
	\subfigure {
		\begin{minipage}{0.175\textwidth}
			\centering
			\includegraphics[width=0.95\columnwidth]{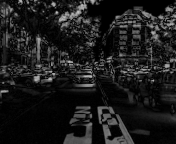}
		\end{minipage}
	}
	\subfigure {
		\begin{minipage}{0.175\textwidth}
			\centering
			\includegraphics[width=0.95\columnwidth]{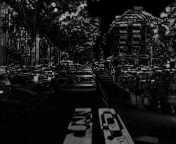}
		\end{minipage}
	}	
	\subfigure {
		\begin{minipage}{0.175\textwidth}
			\centering
			\includegraphics[width=0.95\columnwidth]{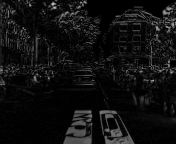}
		\end{minipage}
	}	
	\subfigure {
		\begin{minipage}{0.175\textwidth}
			\centering
			\includegraphics[width=0.95\columnwidth]{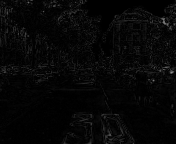}
		\end{minipage}
	}
	\\
	\vspace{0.5cm}
	\subfigure{
		\begin{minipage}{0.175\textwidth}
			\centering
			\includegraphics[width=0.95\columnwidth]{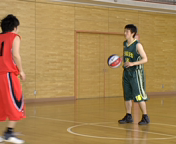}
		\end{minipage}
	}
	\subfigure {
		\begin{minipage}{0.175\textwidth}
			\centering
			\includegraphics[width=0.95\columnwidth]{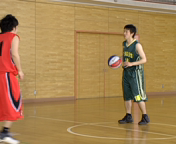}
		\end{minipage}
	}
	\subfigure {
		\begin{minipage}{0.175\textwidth}
			\centering
			\includegraphics[width=0.95\columnwidth]{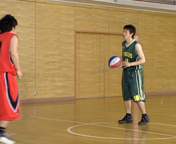}
		\end{minipage}
	}	
	\subfigure {
		\begin{minipage}{0.175\textwidth}
			\centering
			\includegraphics[width=0.95\columnwidth]{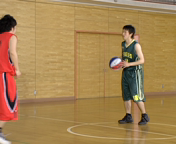}
		\end{minipage}
	}	
	\subfigure {
		\begin{minipage}{0.175\textwidth}
			\centering
			\includegraphics[width=0.95\columnwidth]{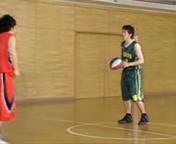}
		\end{minipage}
	}
	\\
	\subfigure{
		\begin{minipage}{0.175\textwidth}
			\centering
			\includegraphics[width=0.95\columnwidth]{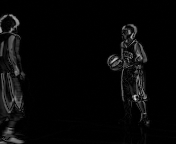}
		\end{minipage}
	}
	\subfigure {
		\begin{minipage}{0.175\textwidth}
			\centering
			\includegraphics[width=0.95\columnwidth]{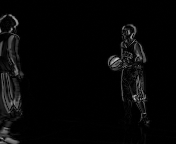}
		\end{minipage}
	}
	\subfigure {
		\begin{minipage}{0.175\textwidth}
			\centering
			\includegraphics[width=0.95\columnwidth]{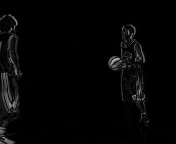}
		\end{minipage}
	}	
	\subfigure {
		\begin{minipage}{0.175\textwidth}
			\centering
			\includegraphics[width=0.95\columnwidth]{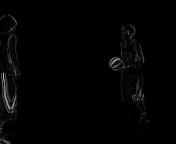}
		\end{minipage}
	}	
	\subfigure {
		\begin{minipage}{0.175\textwidth}
			\centering
			\includegraphics[width=0.95\columnwidth]{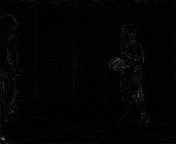}
		\end{minipage}
	}
	\caption{Examples: Kitti 3 (top) and Basketball Drive (bottom). From left to right: $t_{-4}$, $t_{-3}$, $t_{-2}$, $t_{-1}$, $t_0$ (artificial)}
	\label{fig:examples}
\end{figure*}

\begin{table}
	\caption{Mean MSE and SSIM of the different reference pictures at times $t_i$ for all test sequences (unseen during training) with respect to the original picture at time $t_0$.}
	\centering
	\label{tab:kittis}
	\begin{tabular}{cll}
		Reference picture at time & MSE & SSIM \\ \hline
		$t_{-4}$ (conventional) & 2547   &  0.42    \\ 
		$t_{-3}$ (conventional) &  2170  &  0.45    \\
		$t_{-2}$ (conventional) &  1680   &  0.49    \\
		$t_{-1}$ (conventional) &  1033  &  0.60    \\
		$t_{0}$ \hspace{0.2cm} (artificial)\hspace{0.2cm} &  237   &  0.83    \\ 
	\end{tabular}
\end{table}

\begin{table}
	\caption{BD-rate gains and coding time ratios for all videos and mean values. Positive BD-rate gains indicate increased coding efficiency. Coding time ratio $>1$ indicate increased complexity.}
	\centering
	%\small
	\label{tab:overallbd}
	\setlength{\tabcolsep}{0.3em}
	\begin{tabular}{l|cccc|cc}
		
		& \multicolumn{4}{c|}{BD-rates} & \multicolumn{2}{c}{Time ratios}   \\ 
		Video          &  Y     & Cb     &  Cr      &  Weighted &  Enc. & Dec.   \\ \hline
		KITTI 1        & 1.48\% & 5.52\% & 1.50\%   & 1.98\% & 0.63 & 7.36  \\ 
		KITTI 2        & 2.09\% & 7.27\% & -5.82\%  & 1.75\% & 0.71 & 6.86  \\ 
		KITTI 3        & 2.52\% & 4.81\% & -3.42\%  & 2.06\% & 0.69 & 8.02  \\ 
		KITTI 4        & 2.45\% & 5.76\% & -2.40\%  & 2.26\% & 0.75 & 5.27  \\ 
		KITTI 5        & 0.65\% & 3.73\% & 3.85\%   & 1.44\% & 0.66 & 7.94  \\ 
		KITTI 6        & 2.03\% & 3.28\% & -3.31\%  & 1.52\% & 0.70 & 7.61  \\ 
		KITTI 7        & 0.50\% & 3.41\% & 0.25\%   & 0.84\% & 0.69 & 7.64  \\ 
		KITTI 8        & 1.35\% & 3.71\% & -1.06\%  & 1.35\% & 0.73 & 6.81  \\ 
		KITTI 9        & 0.46\% & 3.87\% & -1.08\%  & 0.69\% & 0.67 & 7.43  \\ \hline
		Mean           & 1.50\% & 4.60\% & -1.27\%  & 1.54\% & 0.69 & 7.21 
	\end{tabular}
\end{table}

In this section, we discuss the results of the neural network and of our complete pipeline using our implementation in the HEVC reference software HM~16.18.
The encoder was configured in a low-delay configuration where the four preceding pictures were used as reference pictures.
The results presented in the section were achieved using only sequences which were not used for the training of the neural network.
We chose these sequences as they are best for the demonstration of the network's capability which depends on the ascertainability of the motion.
The motion in the KITTI dataset is better ascertainable then the one of more general MPEG test sequences.

First, we will analyze the generated artificial reference pictures. 
It is difficult to measure the quality of reference pictures with a metric because their performance is revealed only during motion-compensated prediction. 
In consequence, metrics like MSE are limited for making conclusions in this case. 
For example, when comparing two similar pictures where one is translated by a single pel the MSE will not be negligible even though the original picture could be reconstructed nearly perfectly from the translated picture using motion-compensated prediction.
Still, a tendency can be obtained by measuring the quality of the artificial reference pictures in terms of metrics without the context of motion-compensated prediction.
In the first and third row, Fig.~\ref{fig:examples}  shows the four reference pictures for a picture at time $t_0$. The first four pictures are the conventional reference pictures at time instances $t_{-1}$ to $t_{-4}$ and the fifth picture is the generated artificial reference picture at time $t_0$. 
In the second and fourth row, the error between the corresponding picture in the first row and the picture to be coded at time $t_0$ is visualized. The error pictures were generated by calculating the absolute difference between the pictures, thus the whiter a pixel, the higher the error at that point.

Two main observations can be made here. 
Firstly, the error of the artificial reference picture is lower than the error of the conventional reference pictures for both examples. 
Secondly, the error increases with increasing temporal distance between the conventional reference pictures and the picture to be coded at time $t_0$, as expected.
The observations from the two representative examples are the same for a larger dataset. 
We calculated the average MSE and SSIM for 684 snippets unseen during training. 
The results for each of the reference pictures are presented in Table~\ref{tab:kittis}.
It is acknowledged that our method could likely be further improved by deciding which reference picture to replace for every picture to be coded adaptively depending on an analysis of the conventional reference pictures.

The coding efficiency results are summarized in Table~\ref{tab:overallbd}.
BD rates were calculated following \cite{Bjontegaard2008}.
Additionally, as suggested in \cite{Sullivan2011}, weighted average BD rates $\text{BD}_{\text{YCbCr}}$ were calculated with weighting factors of 6/1/1 for the three components Y/Cb/Cr, respectively.
In average, weighted BD-rate gains of 1.54\% were achieved with values up to 2.26\%.
Preliminary results suggest that the neural network can also predict videos of higher resolution.
To get further insights, we also tested our method on completely different sequences (namely MPEG test sequences) whose characteristics vary considerably from the sequences used for training. 
As expected, the neural network does not perform satisfactory enough for those sequences to improve the coding efficiency.
Nevertheless, the previously described example from Fig.~\ref{fig:examples} indicates that this limitation can be overcome.

We evaluated the complexity of our method by measuring the coding time ratios relative to the unmodified HM implementation.
The results are summarized in Tab.~\ref{tab:overallbd}.
The processing time for the neural network was included for the measured times.
The encoder complexity is reduced (69\% of the original time) because the motion estimation of HM speeds up more due to the higher similarity of the artificial reference picture and the original picture than the forward pass of the neural network takes in turn.
On the other hand, the decoder complexity is increased by a factor of 7.2.
This is due to the fact that the decoder needs to perform the forward pass for the neural network but does not benefit from the sped up of the encoder-only motion estimation.

\section{Conclusions}
\label{sec:conclusions}
In this paper, we proposed the generation of artificial reference pictures using deep recurrent neural networks.
The method is based on processing conventional reference picture to create a prediction of an artificial reference picture at the time instance of the currently coded picture.
Thereby, we are able to increase the coding efficiency of HEVC with average BD-rate gains of 1.54\%.

\bibliographystyle{IEEEbib}
\bibliography{references}

\end{document}